\documentclass[a4paper,11pt]{article}
\usepackage{pos}
\usepackage{physics}
\usepackage{float}
\usepackage{subcaption}
\usepackage{amsmath}
\usepackage{dsfont}

\usepackage{tikz}
\usepackage{tabularx}
\usepackage{tikz-cd}

\title{Investigating the variance increase of readout error mitigation through classical bit-flip correction on IBM and Rigetti quantum computers}
\ShortTitle{Investigating the variance increase of readout error mitigation}

\author[a,b]{Constantia Alexandrou}
\author[c,d]{Lena Funcke}
\author[e,b]{Tobias Hartung}
\author[f]{Karl Jansen}
\author[b]{Stefan Kühn}
\author*[a,b]{Georgios Polykratis}
\author[f]{Paolo Stornati}
\author[g]{Xiaoyang Wang}
\author[h]{Tom Weber}
\affiliation[a]{Department of Physics, University of Cyprus, P.O. Box 20537, 1678 Nicosia, Cyprus}
\affiliation[b]{Computation-Based Science and Technology Research Center,
  The Cyprus Institute, 20 Kavafi Street, 2121 Nicosia, Cyprus}
\affiliation[c]{Center for Theoretical Physics, Co-Design Center for Quantum Advantage, and NSF AI Institute for Artificial Intelligence and Fundamental Interactions, Massachusetts Institute of Technology, 77 Massachusetts Avenue, Cambridge, MA 02139, USA}
\affiliation[d]{Perimeter Institute for Theoretical Physics, 31 Caroline Street North, Waterloo, ON N2L 2Y5, Canada}
\affiliation[e]{Department of Mathematical Sciences University of Bath, Bath, United Kingdom}
\affiliation[f]{Deutsches Elektronen-Synchrotron DESY, Platanenallee 6, 15738 Zeuthen, Germany}
\affiliation[g]{School of Physics, Peking University, 5 Yiheyuan Rd, Haidian District, Beijing 100871, China}
\affiliation[h]{Department of Computer Science, Universität Hamburg, Vogt-Kölln-Str.\ 30, 22527 Hamburg, Germany}

\emailAdd{alexand@ucy.ac.cy}  
\emailAdd{lfuncke@mit.edu}
\emailAdd{tobias.hartung@desy.de}
\emailAdd{karl.jansen@desy.de}
\emailAdd{s.kuehn@cyi.ac.cy}
\emailAdd{g.polykratis@cyi.ac.cy}
\emailAdd{paolo.stornati@desy.de}
\emailAdd{zzwxy@pku.edu.cn}
\emailAdd{tom.weber-1@uni-hamburg.de}

\abstract{Readout errors are among the most dominant errors on current noisy intermediate-scale quantum devices. Recently, an efficient and scaleable method for mitigating such errors has been developed, based on classical bit-flip correction~\cite{Funcke:2020olv,Funcke:2021aps}. In this talk, we compare the performance of this method for IBM's and Rigetti's quantum devices, demonstrating how the method improves the noisy measurements of observables obtained on the quantum hardware.
Moreover, we examine the variance amplification to the data after applying of our mitigation procedure, which is common to all mitigation strategies. We derive a new expression for the variance of the mitigated Pauli operators in terms of the corrected expectation values and the noisy variances.
Our hardware results show good agreement with the theoretical prediction, and we demonstrate that the increase of the variance due to the mitigation procedure is only moderate.\\

Preprint number: MIT-CTP/5351} 

\FullConference{%
 The 38th International Symposium on Lattice Field Theory, LATTICE2021
  26th-30th July, 2021
  Zoom/Gather@Massachusetts Institute of Technology
}

\begin{document}
\maketitle

\section{Introduction}

State-of-the-art Markov chain Monte Carlo (MCMC) methods for lattice field theories cease to work in certain parameter regimes due to the infamous sign problem~\cite{PhysRevLett.94.170201}. Prominent examples are QCD in the presence of a baryon chemical potential or a topological term \cite{Fukushima2010}, where the latter is linked to the strong CP problem. In addition, the MCMC approach relies on a Wick rotation of the original theory, resulting in a formulation in Euclidean space-time. Thus, real-time phenomena, such as the out-of-equilibrium dynamics following heavy-ion collisions, are inaccessible with MCMC approaches. As a result, many non-perturbative phenomena cannot be addressed with conventional MCMC techniques. Quantum computing offers to bypass these problems, as it is free from purely numerical limitations and allows for simulating real-time dynamics. First proof-of-principle experiments have already successfully demonstrated these capabilities in lower dimensions~(see, e.g., Refs.~\cite{Martinez2016,Kokail2018,Klco2018,Klco2019,Ciavarella2021,Zhou:2021kdl}). In the long run, quantum computing is therefore one of the most promising new methods for studying unexplored regimes of the Standard Model.

Current noisy intermediate-scale quantum (NISQ)~\cite{Preskill2018quantumcomputingin} computers suffer from several sources of noise, among them gate errors, depolarizing noise, and measurement errors. Measurement errors occur if a measurement result $0$ is misidentified as $1$ or vice versa. They can be among the most common errors on current NISQ computers with error rates reaching up to $\mathcal{O}(10\%)$~\cite{10.1145/3352460.3358265}. While such NISQ devices do not allow for full quantum error correction, the effect of errors can be partially reduced using error mitigation techniques~(see, e.g., Refs.~\cite{Kandala2017,Endo2018,Geller2021}). These techniques typically come at the expense of increasing the variance of the measurement results, thus requiring a larger number of measurements to obtain the same accuracy~\cite{Endo2018}.

In this work, we investigate this effect for a recently developed mitigation protocol for measurement errors~\cite{Funcke:2020olv,Funcke:2021aps} on IBM's and Rigetti's quantum devices. In particular, we compare the change in variance from the hardware results with and without applying the mitigation procedure to the theoretically predicted values and benchmark the performance of the protocol. We also derive a new expression for the variance of the mitigated Pauli expectation values in terms of noisy variances and corrected expectation values, which explicitly excludes the possibility that the variance increases exponentially. This important result, together with the polynomial overhead costs for the mitigation method derived in Ref.~\cite{Funcke:2020olv}, proves the scalability of the method.

\section{Readout Error Mitigation}
\label{sec:readouterrormitigation}

In this section, we briefly review the readout mitigation method proposed in Refs.~\cite{Funcke:2020olv,Funcke:2021aps}. We first focus on the impact of this method on the expectation values and afterwards on the variances.

\subsection{Impact of the Mitigation Procedure on the Expectation Values}

Throughout this paper, we consider a quantum device with $Q$ qubits, which performs projective measurements in the computational basis $\{\ket{0},\ket{1}\}$. We focus on measurement errors (also referred to as bit-flips or readout errors), which result from misidentifying a measurement outcome $0$ as a $1$  and vice versa. For simplicity, we neglect any other error sources at this stage and discuss them briefly in the results section. Thus, we assume that the quantum device transforms the initial state into a pure state $\ket{\psi}$, such that we can measure the expectation value $\langle\psi|O|\psi\rangle$ of an observable~$O$. With appropriate post-rotations applied to $|\psi\rangle$, we can always consider $O$ to be diagonal in the computational basis~\cite{Peruzzo2014}, i.e., we can write $O$ as a string of $\{\mathds{1},Z\}^Q$. Finally, we can treat bit flips on different qubits as uncorrelated, which is a good approximation for the superconducting hardware devices that we use here~\cite{Qiskit:2021, karl2021proceedings}.

For each qubit $q$, there is a probability $p_{q,0}$ of misidentifying the measurement outcome $0$ as a $1$ and vice versa with probability $p_{q,1}$\footnote{Note that this misidentification can happen either before or after the projective measurement.}. Thus, instead of directly measuring the expectation value of an operator $O$, we rather measure the expectation values of ``noisy'' operators $\tilde{O}$ that are subject to these bit flips. Now, we can write $O$ as a linear combination of these noisy operators, such that the resulting expectation value with respect to the bit-flip probabilities gives the desired outcome. To illustrate this, let us consider the case of the operator $Z_q$ acting on a single qubit~$q$. We can express this operator as a linear combination of noisy operators in the following way~\cite{Funcke:2020olv,Funcke:2021aps}:
\begin{equation}
 Z_q= \frac{1}{\gamma(Z_q)}\mathds{E}(\tilde{Z}_q)-\frac{\gamma(\mathds{1})}{\gamma(Z_q)}I\ ,
 \label{eq:Zq}
\end{equation}
where $\mathds{E}(\tilde{Z}_q)$ is the expectation of the noisy operator $\tilde{Z}_q$ with respect to bit flips, and we defined
\begin{equation*}
\gamma(O_q) =
\begin{cases}
    1-p_{q,0}-p_{q,1} & \text{ for } O_q=Z_q\\
    p_{q,0}-p_{q,1} & \text{ for } O_q = \mathds{1}_q.
  \end{cases}
\end{equation*}
Note that ``expectation'' means the expected
value for the noisy operator $\tilde{O}$ subject to bit flips, which is different from its quantum
mechanical expectation value $\bra{\psi}\tilde{O}\ket{\psi}$. For example, for $p_{q,0}=p_{q,1}=0.1$, the operator $\tilde{O}$ will only be correctly implemented (such that $\tilde{O}=O$) with a probability of $(1-p_{q,0})(1-p_{q,1})=0.81$. 

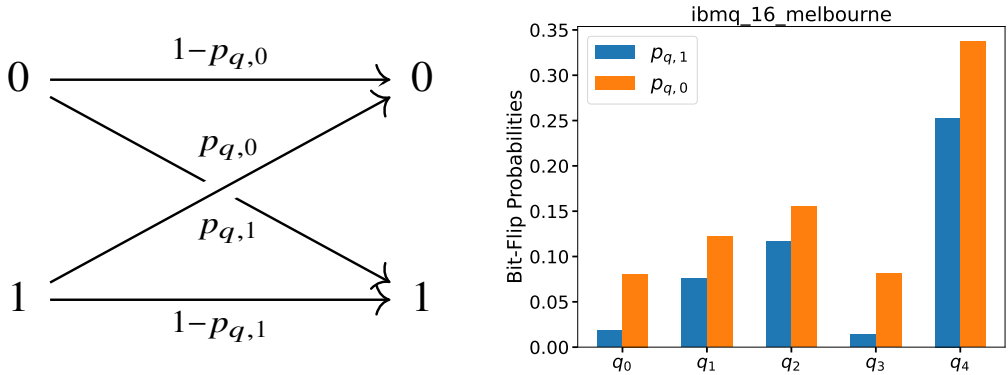
\begin{figure}[htp!]
\centering
\begin{subfigure}[c]{0.45\linewidth}
\begin{tikzpicture}[baseline= (a).base]
\node[scale=1.5] (a) at (0,0){
\begin{tikzcd}[column sep=3cm, row sep=huge]
0 \arrow[r, "1-p_{q,0}"] \arrow[dr, "p_{q,0}", pos=0.4] & 0 \\
1 \arrow[r, "1-p_{q,1}",pos=0.5,swap] \arrow[ur, "p_{q,1}", pos=0.4,swap,crossing over] & 1
\end{tikzcd}
};
\end{tikzpicture}
\end{subfigure}
\begin{subfigure}[c]{0.45\linewidth}
\includegraphics[width=\textwidth]{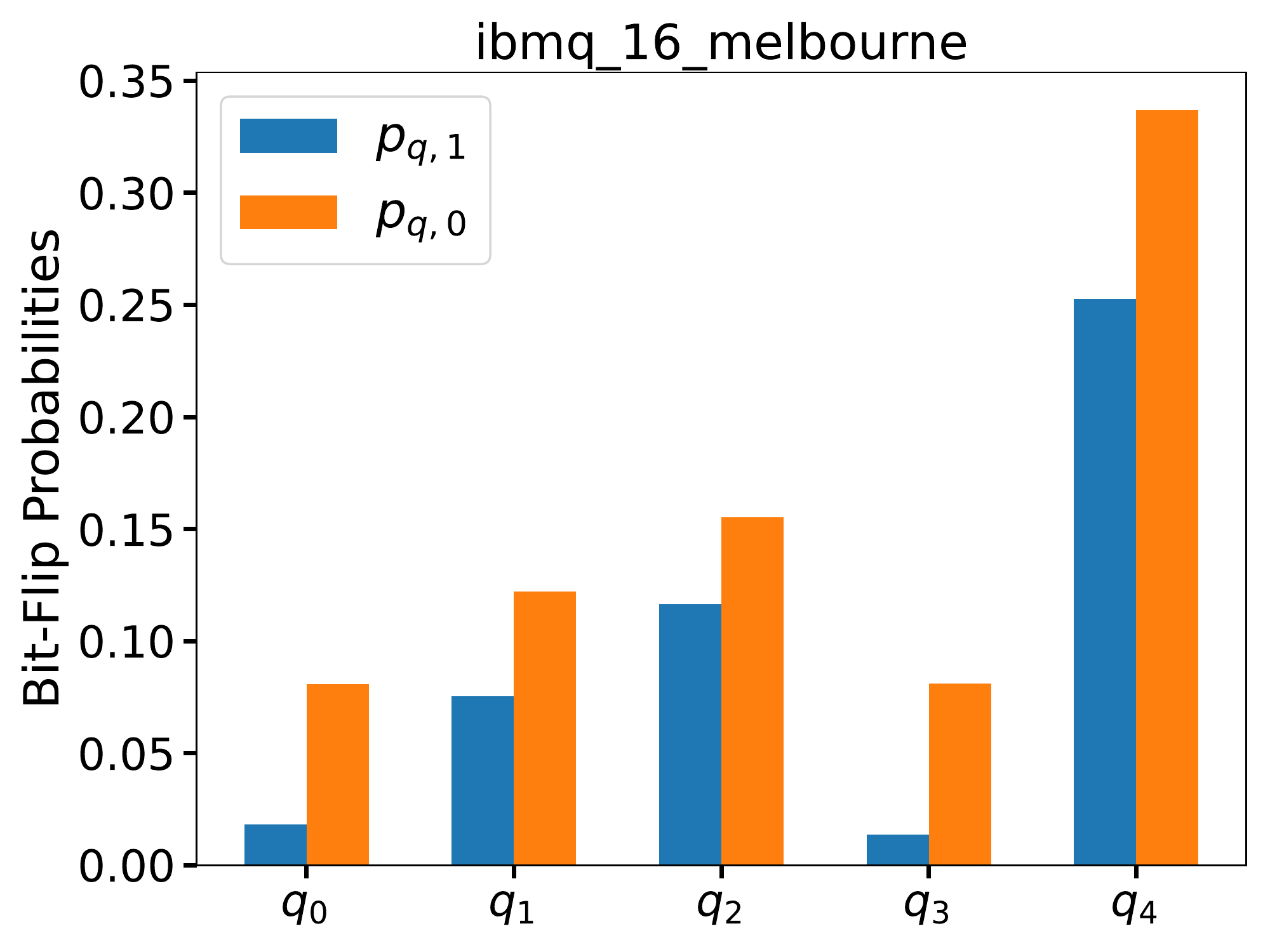}
\end{subfigure}
\caption{Left figure: measurement outcomes (left) and their correct or incorrect identification (right) for a single qubit $q$, where $p_{q,b}$ is the probability of misidentifying the qubit in the state $\ket{b}$ as $\ket{\neg b}$. Right figure: indicative bit-flip calibration of the ibmq\_16\_melbourne device, where $q=q_i$ and $i=0,...,4$.}
\label{fig:my_label}
\end{figure}

To calibrate the values of the bit-flip probability $p_{q,0}$ ($p_{q,1}$), we repeatedly prepare the qubit $q$ in the state $\ket{0}$ ($\ket{1}$) and measure how often we record the incorrect outcome 1 (0), see Fig.~\ref{fig:my_label}.

The single-qubit example in Eq.~\eqref{eq:Zq} can be generalized to an arbitrary number of qubits, and an expansion of the noise-free operator $O=O_Q\otimes\cdots\otimes O_1$ can be obtained in terms of the noisy operators of the space $\{\mathds{1},\tilde{Z}\}^{Q}$, which results in~\cite{Funcke:2020olv} 
\begin{equation}
  \left(O\right)_{O\in\{\mathds{1},Z\}^{\otimes Q}}=
  \omega(O|\tilde{O})^{-1}
  \left(\mathbb{E}\tilde{O}\right)
  _{\tilde{O}\in\{\mathds{1},Z\}^{\otimes Q}},
  \label{eq:corr_exp}
\end{equation}
where the bit-flip probabilities are encoded in the matrix
\begin{equation}
\omega(O|\tilde{O})=\prod^{Q}_{q=1}\Gamma(O_q|\tilde{O}_q) \quad \text{with}\quad \Gamma(O_q|\tilde{O}_q)=
\begin{cases}
    \gamma(O_q) & \text{ for } \tilde{O}_q=\tilde{Z}_q\\
    1 & \text{ for } O_q = I_q \text{ and }\tilde{O}_q = \tilde{I}_q\\
    0 & \text{ for } O_q = Z_q \text{ and }\tilde{O}_q = \tilde{I}_q.\\
  \end{cases}
 \label{omega_matrix}
\end{equation}
Note that $\omega(O|\tilde{O})$ is a lower triagonal matrix with full rank, as long as $p_{q,0} + p_{q,1}\neq 1$ for all qubits~$q$. For reasonable bit-flip probabilities, inverting Eq.~\eqref{omega_matrix} allows for obtaining the corrected expectation values in terms of the noisy expectation measurements.

\subsection{Impact of the Mitigation Procedure on the Variances}
\label{subs:theo_variance}
The mitigation procedure leads to a variance amplification for the error-mitigated results, which is typical for error mitigation techniques~\cite{Endo2018}. In this section, we are interested in the resulting number of additional samples that are necessary to correct for this variance amplification.

The expectation value $\bra{\psi}\tilde{O}\ket{\psi}$ of a noisy operator $\tilde{O}$ is obtained by running the quantum circuit preparing $\ket{\psi}$ and performing a projective measurement multiple times. 
We refer to these repetitions as the number of shots $s$. We then produce a histogram with the data of $N$ experiments using $s$ shots. 
The variance of this histogram contains two components, the bit-flip variance $\mathbb{V}_{\rm bf}$ and the quantum mechanical variance $\mathbb{V}_{\rm QM}$. For the example in Eq.~\eqref{eq:Zq}, we get \cite{Funcke:2020olv}
\begin{equation}
\mathbb{V}\bra{\psi}\tilde{Z}_q\ket{\psi} =\frac{1}{s}\mathbb{V}_{\rm bf}\bra{\psi}\tilde{Z}_q\ket{\psi}+
\frac{1}{s}\mathbb{V}_{\rm QM}\bra{\psi}\tilde{Z}_q\ket{\psi}.
\end{equation}
We can express these variances in terms of the expectation values (see Eq.~(62) in Ref.~\cite{Funcke:2020olv})
\begin{align}
\begin{split}
\label{1qubVar}
\mathbb{V}_{\rm bf}\bra{\psi}\tilde{Z}_q\ket{\psi} &= a_1\bra{\psi}Z_q\ket{\psi}^2-2a_2\bra{\psi}Z_q\ket{\psi} +a_3,\\ \mathbb{V}_{\rm QM}\bra{\psi}\tilde{Z}_q\ket{\psi} &= 1-\bra{\psi}\tilde{Z}_q\ket{\psi}^2,
\end{split}
\end{align}
where the noise-free expectation values of $Z_q$ are given by the error-mitigated expectation values in our computations, and we have defined
\begin{align*}
a_1 &= (p_{q,1}+p_{q,0})(1-p_{q,0}-p_{q,1})+2p_{q,0}p_{q,1},\\
a_2 &= (1-p_{q,0}-p_{q,1})(p_{q,1}-p_{q,0}),\\
a_3 &= (p_{q,0}+p_{q,1}-p^2_{q,0}-p^2_{q,1}).
\end{align*}
For the case of a two-qubit operator with $Z_1$ and $Z_2$ acting on uncorrelated qubits, we can construct the bit-flip variance from the bit-flip variances of the single-qubit operators (see Eq.~(64) in Ref.~\cite{Funcke:2020olv}),
\begin{equation}
\mathbb{V}_{\rm bf}\bra{\psi}\tilde{Z}_2\otimes\tilde{Z}_1\ket{\psi}=
\mathbb{V}_{\rm bf}\tilde{Z}_2\otimes\mathbb{V}_{\rm bf}\tilde{Z}_1
+\mathbb{E}\tilde{Z}_2\otimes\mathbb{V}_{\rm bf}\tilde{Z}_1
+\mathbb{V}_{\rm bf}\tilde{Z}_2\otimes\mathbb{E}\tilde{Z}_1.
\label{eq:var_n2_noisy}
\end{equation}
Using Eq.~(\ref{eq:var_n2_noisy}) and taking the variance of Eq.~(24) in Ref.~\cite{Funcke:2020olv}, we can
express the variances of multi-qubit operators in terms of the variances and expectation values of single-qubit operators,
\begin{equation}
\label{n2_mit_variance_v2}
\begin{aligned}
    \mathbb{V}(Z_2\otimes Z_1)
   =\,&\bigg(\frac{1}{\gamma(Z_2)\gamma(Z_1)}\bigg)^2
    \bigg(\mathbb{V}\tilde{Z}_{2}\otimes\mathbb{V}\tilde{Z}_{1}+\mathbb{V}\tilde{Z}_{2}\otimes(\mathbb{E}\tilde{Z}_1)^2+
    (\mathbb{E}\tilde{Z}_2)^2\otimes\mathbb{V}\tilde{Z}_{1}\bigg)\\
    &+\bigg(\frac{\gamma(I_1)}{\gamma(Z_2)\gamma(Z_1)}\bigg)^2\mathbb{V}\tilde{Z}_2
    +\bigg(\frac{\gamma(I_2)}{\gamma(Z_2)\gamma(Z_1)}\bigg)^2\mathbb{V}\tilde{Z}_1,
\end{aligned}
\end{equation}
where we omitted the $\rm bf$-index for brevity. As discussed above, having access to this variance enables us to predict the number of additional experiments that are necessary to achieve the same accuracy of the computational results after the mitigation.

We note that the novelty of this derivation in Eq.~\eqref{n2_mit_variance_v2} is the explicit prediction of the extent to which the variance is increased after applying the readout error mitigation method. Crucially, we observe that higher-order terms generally have a smaller contribution to the variance because of the smaller prefactors. This guarantees that the variance does not increase exponentially. 

We also note that a similar argument can substantially decrease the computational complexity of the error mitigation method itself. In Eq.~\eqref{eq:corr_exp}, each term corresponds to the case where a different number of bits are flipped, and the coefficients of each of these terms are proportional to a product $\prod_{i}\gamma(I_{i})$, where $i\in\left[0,Q\right]$ are the qubits that are bit-flipped.
Since each $\gamma(I_{i})$ factor is proportional to the bit-flip probabilities, the terms in Eq.~\eqref{eq:corr_exp} that correct for an increasing number of bit flips become decreasingly relevant. This observation is particularly important for the case of a large number of qubits, since we can choose to stop the mitigation method at a lower order, thus decreasing the computational complexity without any notable loss in accuracy.

\section{Numerical Results from Quantum Hardware}
In this section, we measure the expectation values of the two-qubit and three-qubit operators $\bra{\psi}Z_2\otimes Z_1\ket{\psi}$ and $\bra{\psi}Z_3\otimes Z_2\otimes Z_1\ket{\psi}$ on quantum devices from IBM and Rigetti. We examine the impact of the mitigation procedure on both the mean value and the variance of these observables.

\subsection{Results for the Expectation Values}
\label{sec:resultsexp}

To benchmark the method described in Sec.~\ref{sec:readouterrormitigation}, we perform 1000 and 1144 experiments with the IBM and Rigetti quantum devices, respectively. For each experiment, we compute the absolute error
\begin{align}
  \left| \bra{\psi}\tilde{O}\ket{\psi}_{\rm measured} - \bra{\psi}O\ket{\psi}_\text{exact} \right|
  \label{eq:absolute_error},
\end{align}
where the first term refers to the measured expectation value of the operator (either the unmitigated results obtained on the quantum hardware or the mitigated results after applying our method), and the second term is the exact expectation value that we compute analytically. 

\begin{figure}[ht!]
\centering
\begin{tikzpicture}
    \node[anchor=south west,inner sep=0] (image) at (0,0) {\includegraphics[width=0.9\textwidth]{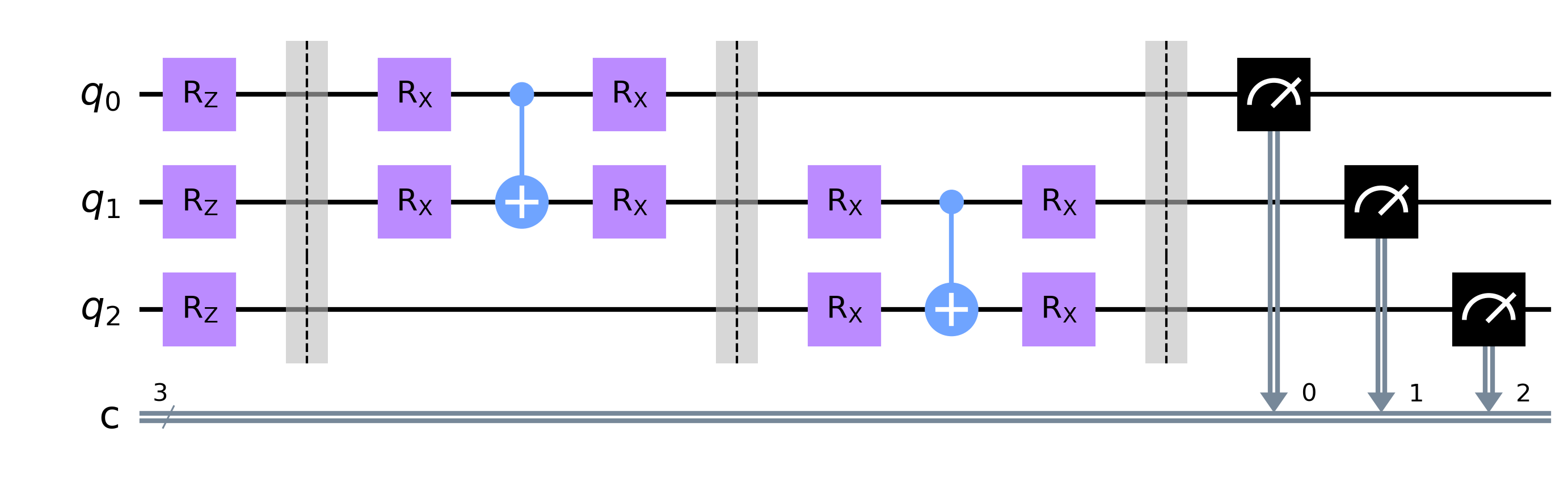}};
    \begin{scope}[x={(image.south east)},y={(image.north west)}]
        \draw[black,dashed,ultra thick,rounded corners] (0.04,0.46) rectangle (0.49,0.95);
    \end{scope}
\end{tikzpicture}
\caption{The quantum circuit used for our experiments with $Q=3$ qubits. For our experiments with $Q=2$ qubits, we used the sub-circuit indicated by the dashed box. The purple $R_X$ and $R_Z$ boxes denote parametric rotation gates, the blue two-qubit connections are CNOT gates, the black boxes are the final measurements, and the vertical dashed lines separate different layers of the quantum circuit.
\label{fig:curcuits}
}
\end{figure}

In our experiments, we choose parametric quantum circuits that are inspired by typical Ansätze used for the variational quantum eigensolver algorithm~\cite{Peruzzo2014}, as shown in Fig.~\ref{fig:curcuits}. Each parameter of the rotation gates is drawn uniformly from $[0,2\pi)$ at the beginning and kept constant over all experiments. In the end, we compute the average of Eq.~\eqref{eq:absolute_error}.
The expectation value $\bra{\psi}\tilde{O}\ket{\psi}_{\rm measured}$
is obtained by preparing $\ket{\psi}$ with the corresponding quantum circuit multiple times and collecting statistics of the measurement outcomes. As before, the number of repetitions is referred to as the number of shots $s$. In case of an ideal, noise-free quantum computer, the average of Eq.~\eqref{eq:absolute_error} should decay as $1/\sqrt{s}$~\cite{Funcke:2020olv,Funcke:2021aps}.  Other errors
on the quantum hardware will lead to a saturation of the average absolute error at a certain value, indicating the level of accuracy that can be reached on the device without 
additional mitigation for these other errors.

As described in Sec.~\ref{sec:readouterrormitigation}, we calibrate the quantum device by preparing a qubit $q$ in a computational basis state $\ket{b}$ and then performing projective measurements. Our estimate of the bit-flip probability $p_{q,b}$ is given by the empirical probability of obtaining the incorrect measurement outcome $\neg b$. Since the initial state of the chosen quantum hardware is the state $\ket{0}$, the states $\ket{b}$ are easily prepared by either applying no gates at all ($\ket{0}$) or applying the corresponding $X$-gates for each qubit ($\ket{1}$). For the case of the IBM hardware, we collect calibration data every time we run a new batch of measurements that uses a different number of shots, in order to have a more accurate description of the noise model. For the case of the Rigetti hardware, we only perform the calibration once because we had exclusive access to the device and thus did not suffer from waiting times.

\begin{figure}[b]
\centering
\begin{subfigure}[c]{.49\linewidth}
\includegraphics[width=0.9\textwidth]{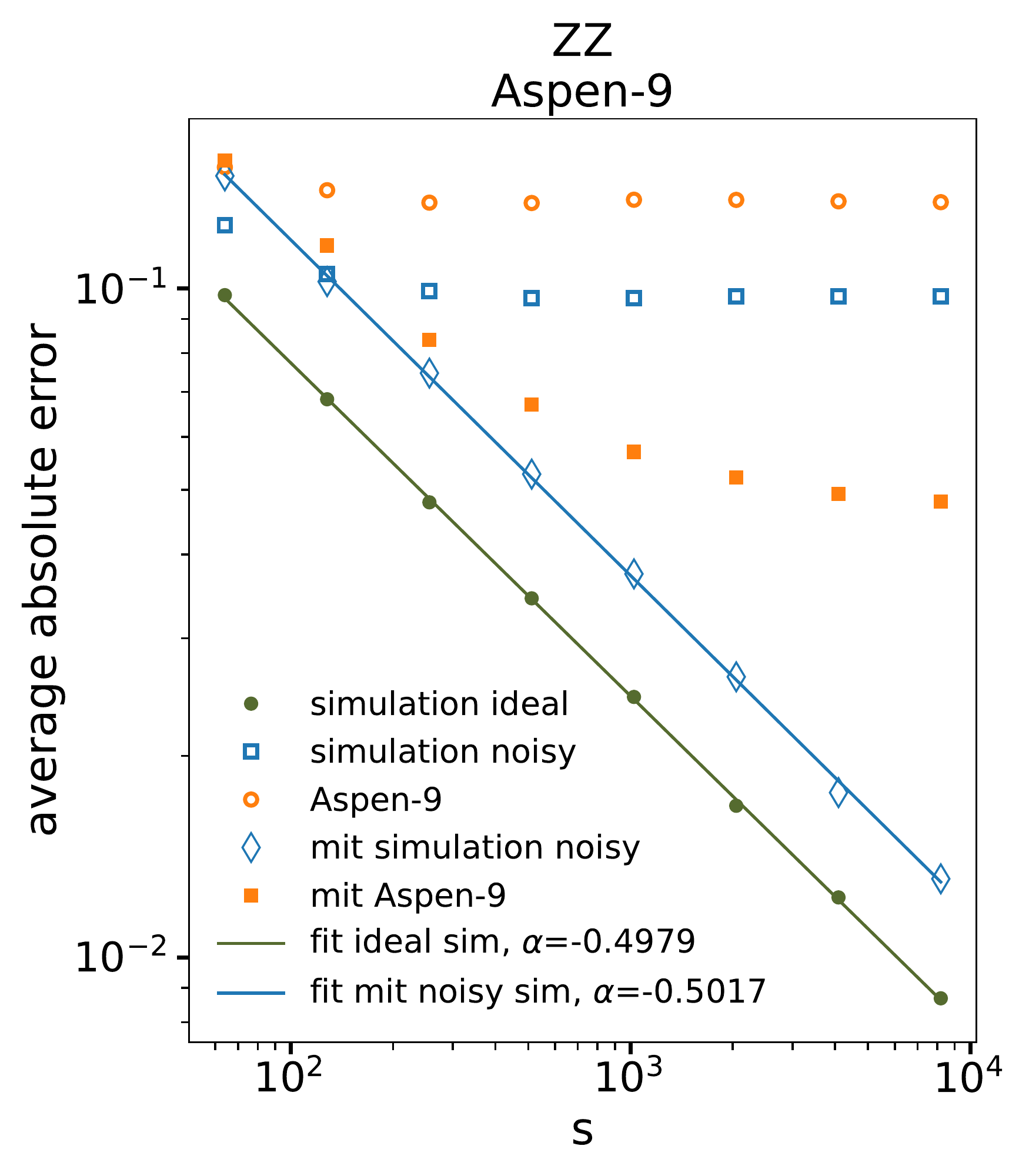}
\end{subfigure}
\begin{subfigure}[c]{.49\linewidth}
\includegraphics[width=0.9\textwidth]{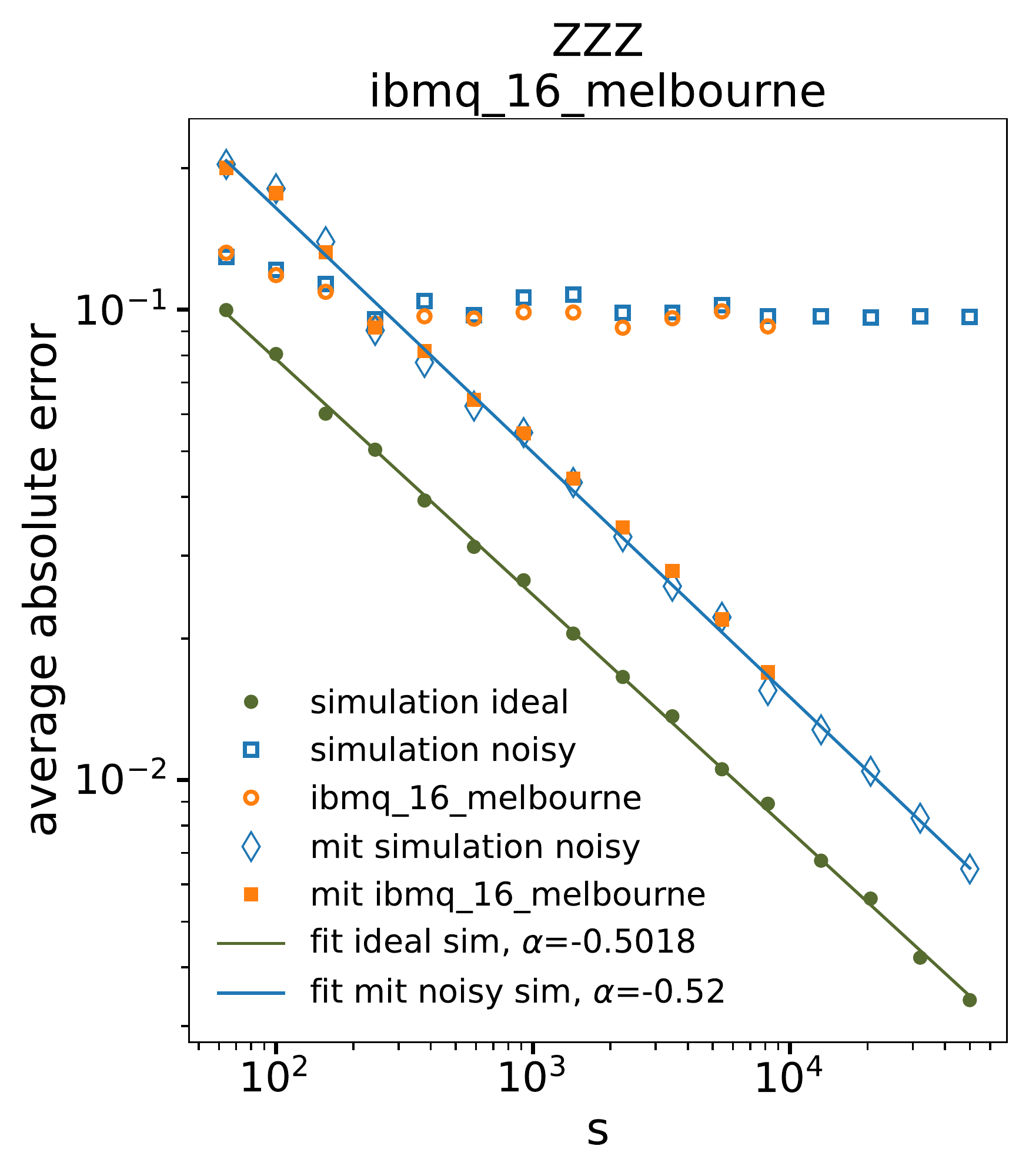}
\end{subfigure}
\caption{Left: average absolute error (see Eq.~\eqref{eq:absolute_error}) of the expectation value $\bra{\psi}Z_2\otimes Z_1\ket{\psi}$ as a function of the number of shots $s$,  obtained on Rigetti's Aspen-9 device. We plot the noisy hardware data with (orange filled squares) and without (orange open circles) error mitigation, as well as a noise-free simulation (green filled circles) and a noisy simulation with (blue open diamonds) and without (blue open squares) error mitigation. Right: same results for $\bra{\psi}Z_3\otimes Z_2\otimes Z_1\ket{\psi}$ obtained on the ibmq\_16\_melbourne device. In both figures, we fit the function $Cs^{\alpha}$ to the ideal (green line) and mitigated (blue line) simulation results, yielding $\alpha\approx -0.5$. The abbreviations \textit{mit} and \textit{sim} refer to the mitigated results and the simulations, respectively.}
\label{fig:same_param}
\end{figure}

Figure~\ref{fig:same_param} shows our results for the average absolute error (see Eq.~\eqref{eq:absolute_error})  as a function of the number of shots $s$. We measure the two-qubit expectation value $\bra{\psi}Z_2\otimes Z_1\ket{\psi}$ on Rigetti's Aspen-9 device (left) and the three-qubit expectation value $\bra{\psi}Z_3\otimes Z_2\otimes Z_1\ket{\psi}$ on the ibmq\_16\_melbourne device (right). We plot the noisy data from the actual quantum devices without error mitigation (orange open circles), as well as the data after applying the measurement error mitigation (orange filled squares). %
We also plot a noise-free simulation (green filled circles) and a noisy simulation with (blue open diamonds) and without (blue open squares) error mitigation. The noise model in the simulation consists only of readout errors.

In Fig.~\ref{fig:same_param}, we fit the function $Cs^{\alpha}$  to the ideal, noise-free simulation results (green line) and the mitigated simulation results (blue line). We obtain $\alpha\approx -0.5$ in both cases and thus see that the errors decay indeed $\propto 1/\sqrt{s}$ with the number of shots $s$, as expected. Without error mitigation, the error of the noisy simulated data initially improves with increasing the number of shots, but eventually saturates at the $\sim 10\%$ level. After applying the mitigation scheme, one again recovers the expected decay $\propto 1/\sqrt{s}$ with the number of shots.

For the data obtained on the quantum hardware, we observe a device-dependent improvement after applying the mitigation method.
On the Aspen-9 device from Rigetti, the mitigation scheme substantially improves the noisy data by reducing the error by half an order of magnitude. At large $s$, we observe a deviation from the expected decay $\propto 1/\sqrt{s}$, because our method can only mitigate the readout error of the noisy data up to a certain accuracy, before the contributions of other underlying noise sources become dominant (see Refs.~\cite{Funcke:2020olv,Funcke:2021aps} for more details).
For the case of the ibmq\_16\_melbourne device, our method reduces the error by almost an order of magnitude. Here, the mitigated noisy results obtained from the quantum device look almost identical to the simulated results, including the expected decay $\propto 1/\sqrt{s}$. Thus, we can attribute the noise almost exclusively to readout errors, which our method successfully corrects for.

\subsection{Results for the Variances}
\begin{figure}[ht!]
\centering
\begin{subfigure}[c]{.45\linewidth}
\includegraphics[width=\linewidth]{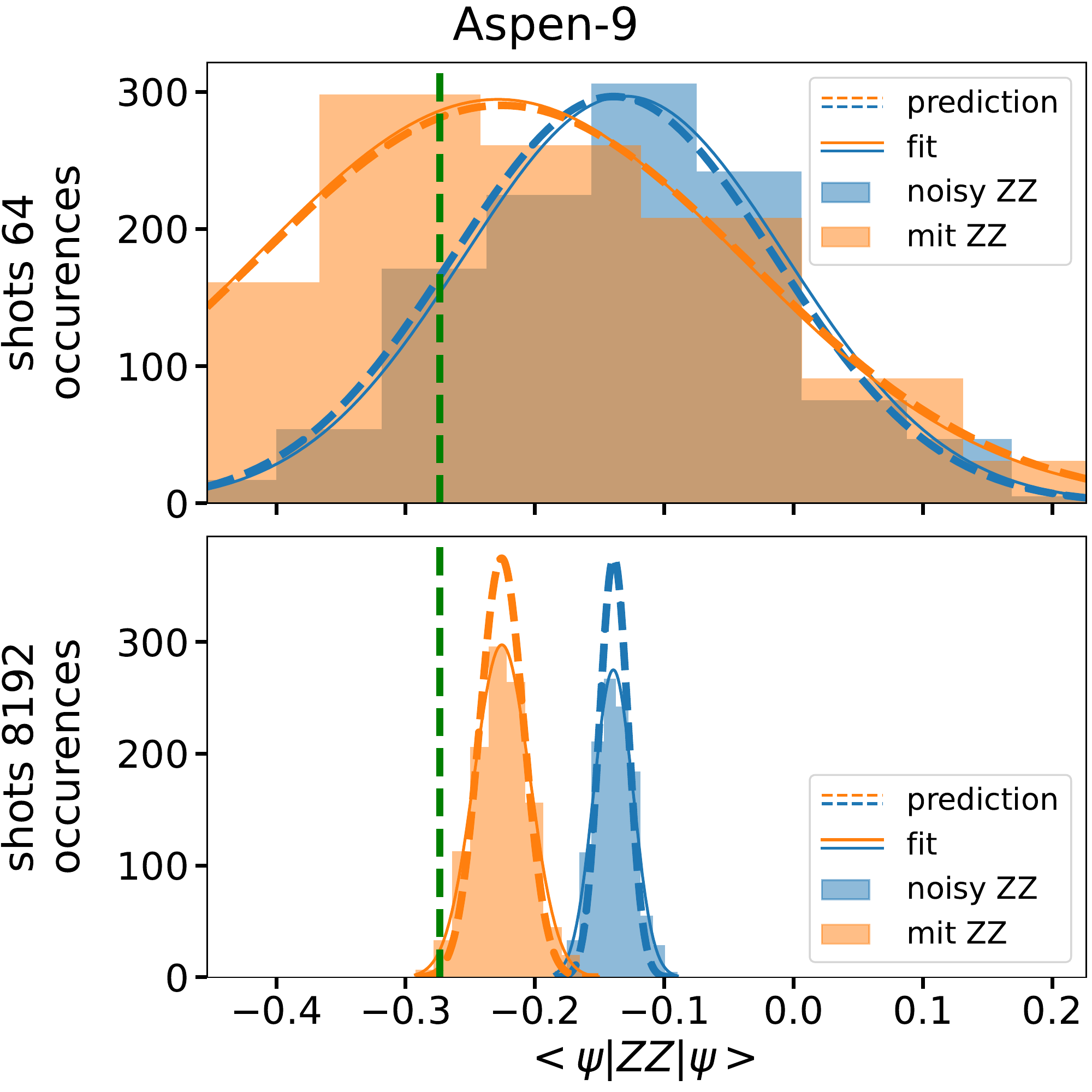}
\end{subfigure}
\begin{subfigure}[c]{.45\linewidth}
\includegraphics[width=\textwidth]{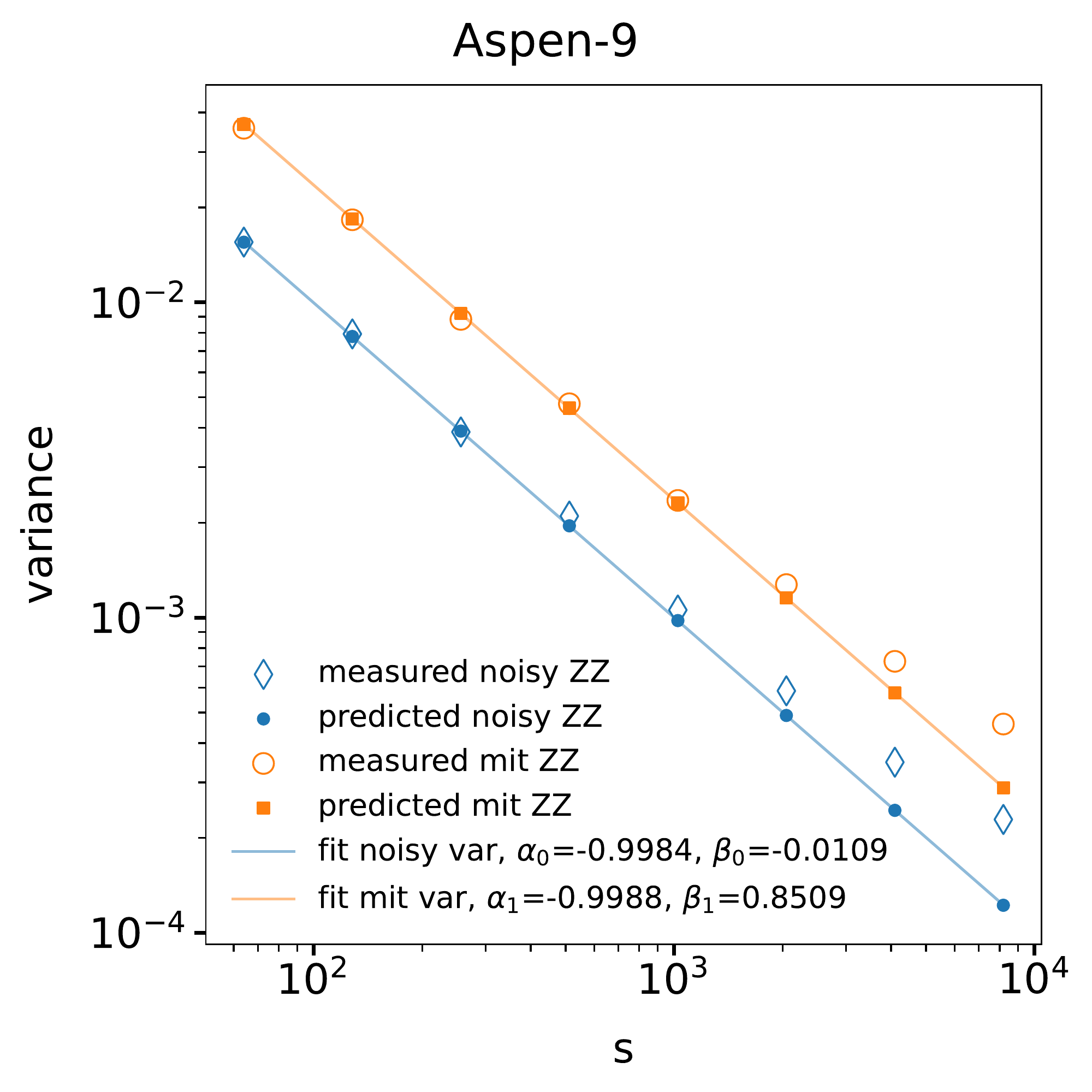}
\end{subfigure}
\caption{Left: Measured distributions of the two-qubit expectation value $\langle\psi|Z_2\otimes Z_1|\psi\rangle$ from the Aspen-9 device of Rigetti, before (blue) and after (orange) the mitigation. Note that the variances are slightly larger in the latter case. We plot the distribution for 64 (upper panel) and 8192 shots (lower panel) to demonstrate the decrease in variance when taking more shots. For a larger number of shots, the measured variances slightly deviate from our prediction, due to other error sources beyond readout errors that we do not mitigate for (see Refs.~\cite{Funcke:2020olv,Funcke:2021aps}). Right: Variances measured on the Aspen-9 quantum device, as a function of the number of shots $s$. We plot the predicted (blue filled points) and measured (blue open diamonds) variances before the mitigation, as well as the predicted (orange filled squares) and measured (orange open circles) variances after the mitigation. We fit the function 
$C_is^{\alpha_i}=e^{\beta_i}s^{\alpha_i}$ to the noisy ($i=0$) and mitigated ($i=1$) measurement results, and use the fit coefficients $\alpha_i$ and $\beta_i$ later in Eq.~\eqref{eq:increase}.
\label{fig:n3_aspen9_same_param}}
\end{figure}

\begin{figure}[ht!]
\centering
\begin{subfigure}[c]{.45\linewidth}
\includegraphics[width=\linewidth]{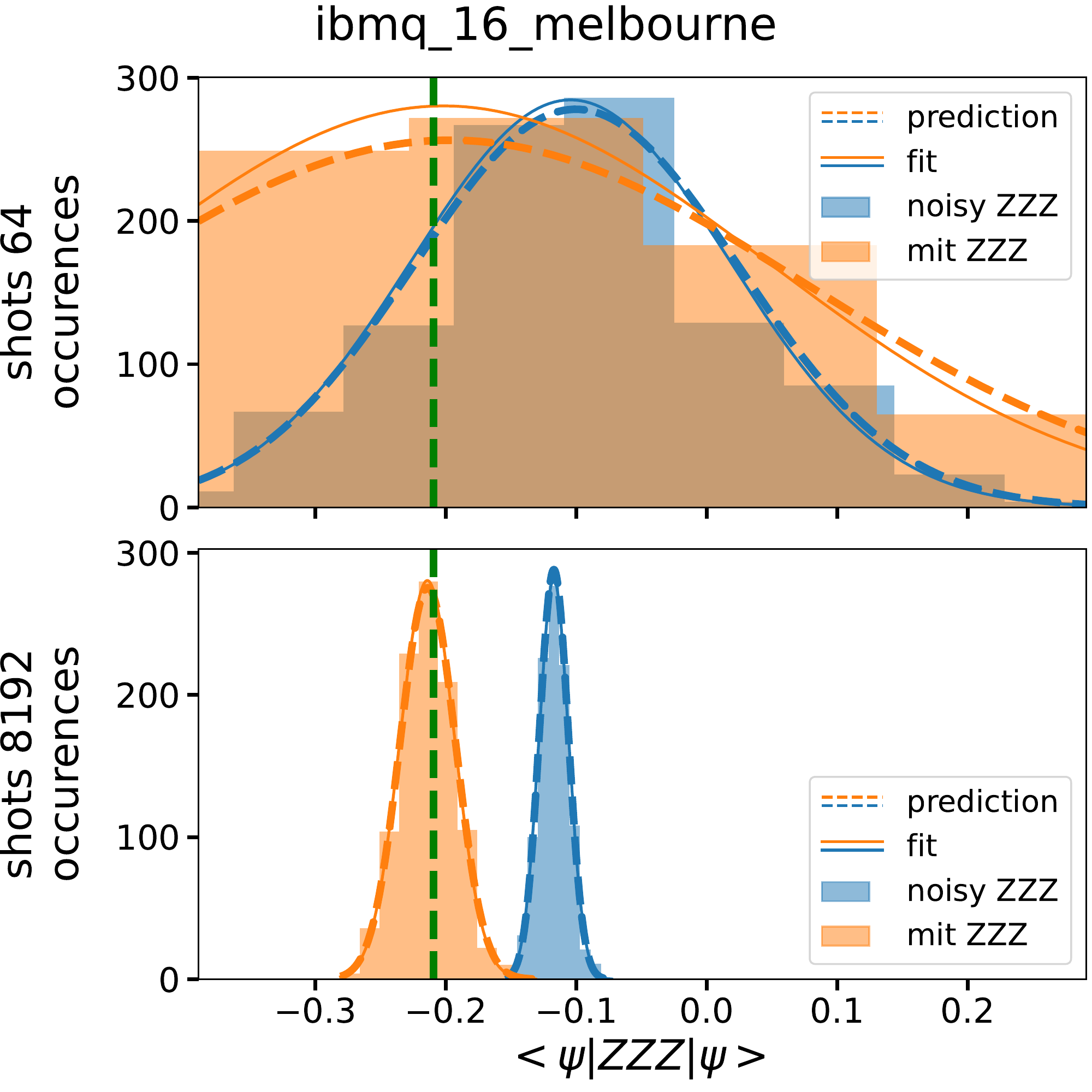}
\end{subfigure}
\begin{subfigure}[c]{.45\linewidth}
\includegraphics[width=\textwidth]{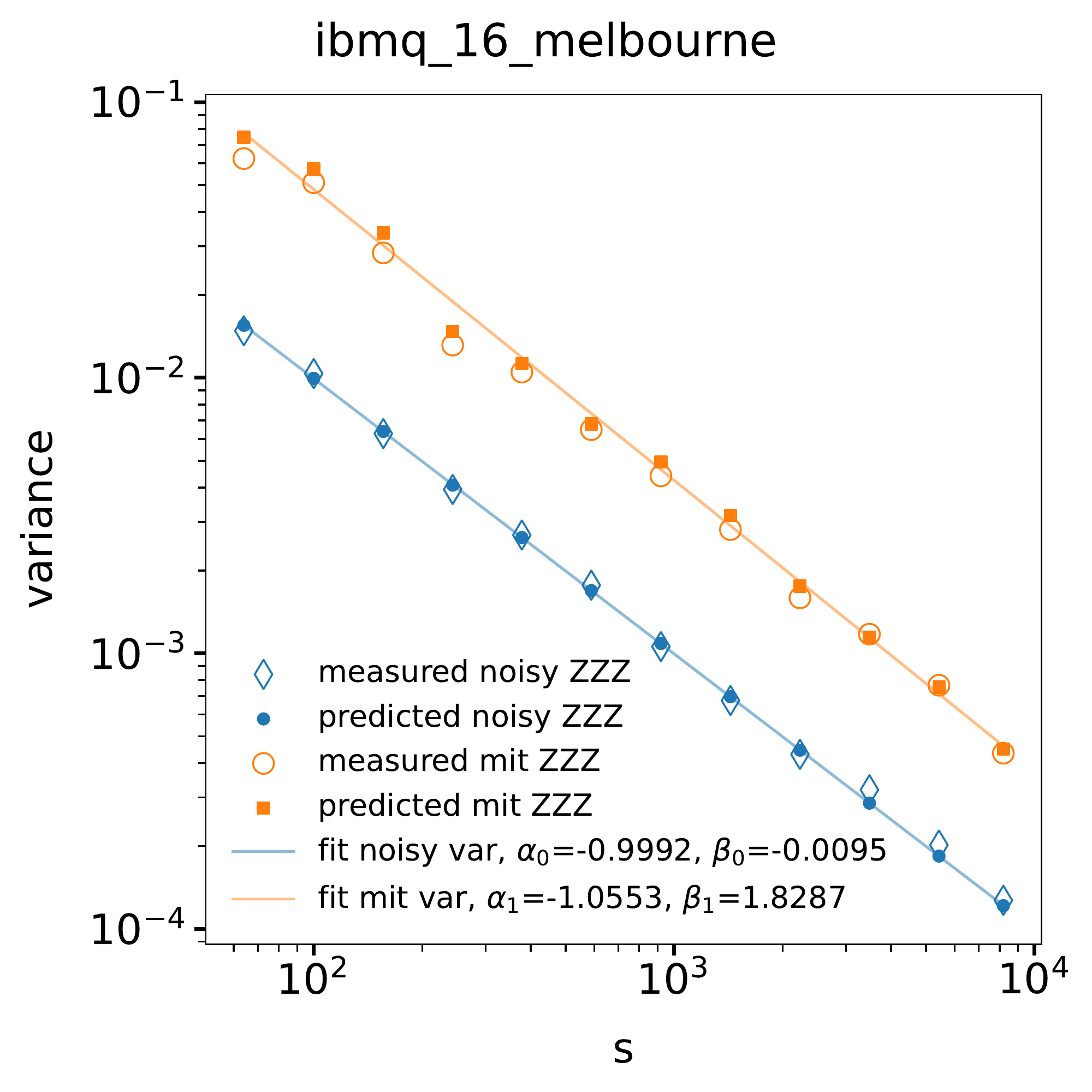}
\end{subfigure}
\caption{Same description as for Fig.~\ref{fig:n3_aspen9_same_param}, with the only difference that we now consider the three-qubit expectation value $\langle\psi|Z_3\otimes Z_2\otimes Z_1|\psi\rangle$ measured on the ibmq\_16\_melbourne device. The measured and predicted variances now agree well for both the small (top left) and the large (bottom left) number of shots.
\label{fig:n3_melbourne_same_param}
}
\end{figure}

In this section, we compare our variance predictions from Sec.~\ref{sec:readouterrormitigation} to the actual variances of the measured distributions for the expectation values $\bra{\psi}Z_2\otimes Z_1\ket{\psi}$ and $\bra{\psi}Z_3\otimes Z_2\otimes Z_1\ket{\psi}$. Figure~\ref{fig:n3_aspen9_same_param} shows the results for the two-qubit experiments performed on the Aspen-9 device of Rigetti. Figure~\ref{fig:n3_melbourne_same_param} shows the results of our three-qubit experiments performed on the ibmq\_16\_melbourne device. 
In the left panels of Fig.~\ref{fig:n3_aspen9_same_param} and Fig.~\ref{fig:n3_melbourne_same_param}, we plot the histograms of the noisy and error-mitigated expectation values for different numbers of shots. In these plots, we show Gaussian fits to the histograms (continuous lines) as well as Gaussian distributions using the predicted variances (dashed lines). In each histogram plot, we also show the true expectation values of the noise-free operators (green vertical lines). In the right panels, we plot the predicted (blue filled points) and measured (blue open diamonds) variances before the mitigation, as well as the predicted (orange filled squares) and measured (orange open circles) variances after the mitigation, as a function of the number of shots $s$.
We also add exponential fits of the form $C_is^{\alpha_i}=e^{\beta_i}s^{\alpha_i}$ to our data and extract the fit coefficients $\alpha_i$ and $\beta_i$, whose indices refer to the noisy ($i=0$) and mitigated ($i=1$) variances, respectively. We will use these extracted coefficients later in Eq.~\eqref{eq:increase}.

As we can see in Fig.~\ref{fig:n3_aspen9_same_param}, the measured variances for the two-qubit Aspen-9 data become larger than the predicted ones as the number of shots increases, both for the noisy and error-mitigated distributions. As discussed in Sec.~\ref{sec:resultsexp}, this deviation is expected due to other types errors beyond the readout error that our mitigation method corrects for.
In Fig.~\ref{fig:n3_melbourne_same_param}, we see that for the three-qubit results of the ibmq\_16\_melbourne device, the predictions for the variance agree well for both smaller and larger number of shots, both for the case of noisy and mitigated results. Here, the mitigation procedure shifts the histogram means to the true expecation value of the noise-free operator. These results agree with the ones obtained in Sec.~\ref{sec:resultsexp}, where we concluded that noise on the ibmq\_16\_melbourne device can be almost exclusively attributes to readout errors.

In both Fig.~\ref{fig:n3_aspen9_same_param} and Fig.~\ref{fig:n3_melbourne_same_param}, one can see that the application of the mitigation scheme slightly increases the variance of the data. As we have already discussed in Sec.~\ref{subs:theo_variance}, this is characteristic for all
error mitigation procedures, and a larger number of samples is needed to achieve the same accuracy~\cite{Endo2018}. We can use our variance predictions to estimate this additional number of samples. If $s_{0}$ is the number of samples without error mitigation and $s_{1}$ is the number of samples when applying error mitigation, then the two variances are the same if $C_0s^{\alpha_0}_0=C_1s^{\alpha_1}_1$. Using this equality and our previous definition of $C_i\equiv e^{\beta_i}$, we derive that we need 
\begin{equation}
\label{eq:increase}
\frac{s_1}{s_0}=\left(\frac{C_0}{C_1}\right)^{\frac{1}{\alpha_1}}s^{\frac{\alpha_0}{\alpha_1}-1}_0
= e^{\frac{\beta_0-\beta_1}{\alpha_1}}\,
s^{\frac{\alpha_0}{\alpha_1}-1}_0
\end{equation}
times more samples to compensate for the variance amplification of our mitigation procedure.
In the case of large statistics, we can use the approximation $\alpha_0\approx\alpha_1$, so the above formula becomes a constant, $\frac{s_1}{s_0} \approx \exp(\frac{\beta_0-\beta_1}{\alpha_1})$.
For the two-qubit implementation on the Aspen-9 device (see Fig.~\ref{fig:n3_aspen9_same_param}), the fit coefficients  indicate that we need $2.4 s^{-0.0004}_0$ times more samples to retain the same accuracy after the mitigation. For the three-qubit implementation on the ibmq\_melbourne device (see Fig.~\ref{fig:n3_melbourne_same_param}), we need $5.7s^{-0.053}_0$ times more samples. Thus,  these overhead costs are moderate in both cases.

\section{Conclusion}

In this work, we benchmarked a recently proposed mitigation scheme for readout errors~\cite{Funcke:2020olv,Funcke:2021aps} on quantum devices from IBM and Rigetti. For both devices, we found a substantial improvement of the computed expectation values. The two-qubit experiments on Rigetti's Aspen-9 machine show an improvement of the average absolute error by half an order of magnitude. The average error for the three-qubit experiments on IBM's ibmq\_16\_melbourne machine was improved by almost an order of magnitude. We also derived and experimentally tested the theoretical predictions for the variance amplification that is caused by applying the mitigation scheme. For a small number of shots, the predictions agree well with the variances of the distributions measured on the quantum hardware. For a large number of shots, the variance for the data obtained on the Rigetti machine slightly deviates from the prediction, while the IBM experiments again show good agreement. In the former case, the deviations are likely caused by other types of errors beyond the readout error that our mitigation method corrects for. In all experiments, only a moderate number of additional samples is required to compensate for the variance amplification and thus to achieve the same accuracy after mitigation. From a theoretical perspective, our key new result is the sub-exponential increase of the variance, which, together with the polynomial scaling of the error mitigation method derived in~Ref.~\cite{Funcke:2020olv}, demonstrates the scalability of the mitigation method.

\acknowledgments
Research at Perimeter Institute is supported in part by the Government of Canada through the Department of Innovation, Science and Industry Canada and by the Province of Ontario through the Ministry of Colleges and Universities.
L.F.\ is partially supported by the U.S.\ Department of Energy, Office of Science, National Quantum Information Science Research Centers, Co-design Center for Quantum Advantage (C$^2$QA) under contract number DE-SC0012704, by the DOE QuantiSED Consortium under subcontract number 675352, by the National Science Foundation under Cooperative Agreement PHY-2019786 (The NSF AI Institute for Artificial Intelligence and Fundamental Interactions, http://iaifi.org/), and by the U.S.\ Department of Energy, Office of Science, Office of Nuclear Physics under grant contract numbers DE-SC0011090 and DE-SC0021006.
S.K.\ acknowledges financial support from the Cyprus Research and Innovation Foundation under project ``Future-proofing Scientific Applications for the Supercomputers of Tomorrow (FAST)'', contract no.\ COMPLEMENTARY/0916/0048. 
G. P.\ is financially supported by the Cyprus  Research and Innovation Foundation under contract number POST-DOC/0718/0100 and from project NextQCD, co-funded by the European Regional Development Fund and the Republic of Cyprus through the Research and Innovation Foundation with contract id EXCELLENCE/0918/0129. T.W.\ acknowledges the support by DASHH (Data Science in Hamburg - HELMHOLTZ Graduate School for the Structure of Matter) with the Grant-No.\ HIDSS-0002. We would like to thank Rigetti Computing for providing exclusive access to their Aspen-9 quantum device and acknowledge the use of IBM Quantum services for this work. The views expressed are those of the authors, and do not reflect the official policy or position of Rigetti Computing, IBM, or the IBM Quantum team.

\bibliographystyle{JHEP}
\bibliography{bibliography}

\providecommand{\href}[2]{#2}\begingroup\raggedright\begin{thebibliography}{10}

\bibitem{Funcke:2020olv}
L.~Funcke, T.~Hartung, K.~Jansen, S.~K\"uhn, P.~Stornati and X.~Wang,
  \emph{{Measurement Error Mitigation in Quantum Computers Through Classical
  Bit-Flip Correction}},  \href{https://arxiv.org/abs/2007.03663}{{\ttfamily
  2007.03663}}.

\bibitem{Funcke:2021aps}
L.~Funcke, T.~Hartung, K.~Jansen, S.~K\"uhn, M.~Schneider, P.~Stornati et~al.,
  \emph{{Towards Quantum Simulations in Particle Physics and Beyond on Noisy
  Intermediate-Scale Quantum Devices}},
  \href{https://arxiv.org/abs/2110.03809}{{\ttfamily 2110.03809}}.

\bibitem{PhysRevLett.94.170201}
M.~Troyer and U.-J.~Wiese, \emph{Computational complexity and fundamental
  limitations to fermionic quantum monte carlo simulations},
  \href{https://doi.org/10.1103/PhysRevLett.94.170201}{\emph{Phys. Rev. Lett.}
  {\bfseries 94} (2005) 170201}.

\bibitem{Fukushima2010}
K.~Fukushima and T.~Hatsuda, \emph{{The phase diagram of dense QCD}},
  \href{https://doi.org/10.1088/0034-4885/74/1/014001}{\emph{Rept. Prog. Phys.}
  {\bfseries 74} (2011) 014001}.

\bibitem{Martinez2016}
E.A.~Martinez, C.A.~Muschik, P.~Schindler, D.~Nigg, A.~Erhard, M.~Heyl et~al.,
  \emph{Real-time dynamics of lattice gauge theories with a few-qubit quantum
  computer}, \href{https://doi.org/10.1038/nature18318}{\emph{Nature}
  {\bfseries 534} (2016) 516}.

\bibitem{Kokail2018}
C.~Kokail, C.~Maier, R.~van Bijnen, T.~Brydges, M.K.~Joshi, P.~Jurcevic et~al.,
  \emph{{Self-Verifying Variational Quantum Simulation of the Lattice Schwinger
  Model}}, \href{https://doi.org/10.1038/s41586-019-1177-4}{\emph{Nature}
  {\bfseries 569} (2019) 355}.

\bibitem{Klco2018}
N.~Klco, E.F.~Dumitrescu, A.J.~McCaskey, T.D.~Morris, R.C.~Pooser, M.~Sanz
  et~al., \emph{{Quantum-classical computation of Schwinger model dynamics
  using quantum computers}},
  \href{https://doi.org/10.1103/physreva.98.032331}{\emph{Phys. Rev. A}
  {\bfseries 98} (2018) }.

\bibitem{Klco2019}
N.~Klco, J.R.~Stryker and M.J.~Savage, \emph{{SU(2) non-Abelian gauge field
  theory in one dimension on digital quantum computers}},
  \href{https://doi.org/10.1103/PhysRevD.101.074512}{\emph{Phys. Rev. D}
  {\bfseries 101} (2020) 074512}.

\bibitem{Ciavarella2021}
A.~Ciavarella, N.~Klco and M.J.~Savage, \emph{{Trailhead for quantum simulation
  of SU(3) Yang-Mills lattice gauge theory in the local multiplet basis}},
  \href{https://doi.org/10.1103/PhysRevD.103.094501}{\emph{Phys. Rev. D}
  {\bfseries 103} (2021) 094501}.

\bibitem{Zhou:2021kdl}
Z.-Y.~Zhou, G.-X.~Su, J.C.~Halimeh, R.~Ott, H.~Sun, P.~Hauke et~al.,
  \emph{{Thermalization dynamics of a gauge theory on a quantum simulator}},
  \href{https://arxiv.org/abs/2107.13563}{{\ttfamily 2107.13563}}.

\bibitem{Preskill2018quantumcomputingin}
J.~Preskill, \emph{Quantum {C}omputing in the {NISQ} era and beyond},
  \href{https://doi.org/10.22331/q-2018-08-06-79}{\emph{{Quantum}} {\bfseries
  2} (2018) 79}.

\bibitem{10.1145/3352460.3358265}
S.S.~Tannu and M.K.~Qureshi, \emph{Mitigating measurement errors in quantum
  computers by exploiting state-dependent bias},  in \emph{Proceedings of the
  52nd Annual IEEE/ACM International Symposium on Microarchitecture}, MICRO
  '52, (New York, NY, USA), p.~279–290, Association for Computing Machinery,
  2019, \href{https://doi.org/10.1145/3352460.3358265}{DOI}.

\bibitem{Kandala2017}
A.~Kandala, A.~Mezzacapo, K.~Temme, M.~Takita, M.~Brink, J.M.~Chow et~al.,
  \emph{Hardware-efficient variational quantum eigensolver for small molecules
  and quantum magnets},
  \href{https://doi.org/10.1038/nature23879}{\emph{Nature} {\bfseries 549}
  (2017) 242}.

\bibitem{Endo2018}
S.~Endo, S.C.~Benjamin and Y.~Li, \emph{Practical quantum error mitigation for
  near-future applications},
  \href{https://doi.org/10.1103/PhysRevX.8.031027}{\emph{Phys. Rev. X}
  {\bfseries 8} (2018) 031027}.

\bibitem{Geller2021}
M.R.~Geller, \emph{{Conditionally Rigorous Mitigation of Multiqubit Measurement
  Errors}}, \href{https://doi.org/10.1103/PhysRevLett.127.090502}{\emph{Phys.
  Rev. Lett.} {\bfseries 127} (2021) 090502}.

\bibitem{Peruzzo2014}
A.~Peruzzo, J.~McClean, P.~Shadbolt, M.-H.~Yung, X.-Q.~Zhou, P.J.~Love et~al.,
  \emph{A variational eigenvalue solver on a photonic quantum processor},
  \href{https://doi.org/10.1038/ncomms5213}{\emph{Nat. Commun.} {\bfseries 5}
  (2014) 1}.

\bibitem{Qiskit:2021}
{Qiskit Aer API documentation and source code}.

\bibitem{karl2021proceedings}
C.~Alexandrou, L.~Funcke, T.~Hartung, S.~K\"uhn, K.~Jansen, G.~Polykratis
  et~al., \emph{Using classical bit-flip correction for error mitigation in
  quantum computations including 2-qubit correlations},
  \href{https://arxiv.org/abs/2111.08551}{{\ttfamily 2111.08551}}.

\end{thebibliography}\endgroup

\end{document}